\documentclass[preprint,superscriptaddress,aps]{revtex4}

\usepackage{amsmath}
\usepackage {xcolor}
\usepackage{comment}
\usepackage{enumerate}

\usepackage{mathrsfs}
\usepackage{comment}
\usepackage{amssymb}
\usepackage{graphicx}
\usepackage{subfigure}
\usepackage[colorlinks,
            linkcolor=blue,
            anchorcolor=blue,
            citecolor=blue]{hyperref}
\newtheorem{theorem}{Theorem}

\newtheorem{corollary}{Corollary}

\def\ra{\rangle}
\def\la{\langle}

\usepackage{graphicx}
\usepackage{dcolumn}
\usepackage{bm}

\begin{document}

\title{A note on Wigner-Yanase skew information-based uncertainty of quantum channels}

\author{Qing-Hua Zhang}
\email[]{qhzhang@csust.edu.cn}

\affiliation{School of Mathematics and Statistics, Changsha University of Science and Technology, Changsha 410114, China}

\author{Jing-Feng Wu}
\email[]{2210501019@cnu.edu.cn}
\affiliation{School of Mathematical Sciences, Capital Normal University,
Beijing 100048, China}

\author{Shao-Ming Fei}
\email[]{feishm@cnu.edu.cn}
\affiliation{School of Mathematical Sciences, Capital Normal University,
Beijing 100048, China}
\affiliation{Max-Planck-Institute for Mathematics in the Sciences, 04103 Leipzig, Germany}

\begin{abstract}
The variance of quantum channels involving a mixed state gives a hybrid of classical and quantum uncertainties. We seek certain decomposition of variance into classical and quantum parts in terms of the Wigner-Yanase skew information. Generalizing the uncertainty relations for quantum observables to quantum channels, we introduce a new quantity with better quantum mechanical nature to describe the uncertainty relations for quantum channels.
We derive several uncertainty relations for quantum channels via variances and the Wigner-Yanase skew information.
\end{abstract}

\maketitle

\section{Introduction}
Uncertainty principle is considered to be one of the fundamental building blocks of quantum mechanics. The conceptual notion of uncertainty principle was firstly proposed by Heisenberg in 1927, shortly after the emergence of quantum mechanics~\cite{heisenberg1927uber}. The famous Heisenberg-Robertson uncertainty relation indicates the limitation on the precision of simultaneously measuring two observables $X$ and $Y$~\cite{Kennard1927Zur,robertson1929the},
\begin{equation}
\Delta^2 X\Delta^2 Y\geqslant \frac{1}{4}|\la \psi |[X,Y]|\psi\ra|^2,
\end{equation}
where $\Delta^2 \Omega={\la \Omega^2\ra-\la\Omega\ra^2}$ is the variance of an observable $\Omega$ with respect to the measured state $|\psi\ra$, the commutator $[X,Y]=XY-YX$ manifests the characteristic of quantum mechanics. A stronger uncertainty relation was proposed by Schr{\"o}dinger~\cite{schrodinger1930sitzungsberichte}:
\begin{equation}
\Delta^2 X \Delta^2 Y-\left|\operatorname{Re}\left\{\operatorname{Cov}_\rho(X, Y)\right\}\right|^2 \geqslant \frac{1}{4}|\mathbf{tr}(\rho[X, Y])|^2,
\end{equation}
where the covariance is defined by $\operatorname{Cov}_\rho(X, Y) = \mathbf{tr}(\rho XY)-\mathbf{tr}(\rho X)\mathbf{tr}(\rho Y)$ and $\operatorname{Re}$ denotes the real part of a complex number.

The Wigner-Yanase skew information~\cite{wigner1963information,luo2003wigner,Luo2004On}, as the noncommutativity between a quantum state $\rho$ and an observable $X$, is define by $I_\rho (A)=-\frac{1}{2}\mathbf{tr}([\sqrt{\rho},A]^2)$. When the quantum state $\rho$ is a mixed one, the variance is a hybrid of both classical mixing and quantum uncertainty, while the Wigner-Yanase skew information stands for the quantum uncertainty. To catch better quantum mechanical nature, Luo defined the quantity $U_\rho(X)=\sqrt{\Delta^4_\rho(X)-[\Delta^2_\rho(X)-I_\rho(X)]^2}$ to describe the quantum uncertainty~\cite{luo2005heisenberg}. Due to $I_\rho (X) \leqslant U_\rho(X)\leqslant \Delta^2_\rho(X)$, naturally one has~\cite{PhysRevA.82.034101},
\begin{equation}
U_\rho (X) U_\rho (Y)- \left|\operatorname{Re}\left\{\operatorname{Corr}_\rho(X, Y)\right\}\right|^2 \geqslant \frac{1}{4}|\mathbf{tr}(\rho[X, Y])|^2.
\end{equation}

The quantum channel is a fundamental concept in quantum mechanics, which is important in state transformation and information transmitting~\cite{nielsen2002quantum}. The relations between states and channels dictate many aspects of quantum scenarios~\cite{PhysRevLett.106.120401,PhysRevA.77.042303,PhysRevLett.88.017901,busch1997operational,luo2017quantum,PhysRevA.106.012436,PhysRevA.85.032117,PhysRevA.95.042337}. In particular, Luo and Sun~\cite{luo2017quantum} indicated that the coherence and quantum uncertainty are dual viewpoints of the same quantum substrate in terms of skew information of quantum channels. Moreover, in terms of skew information, well-defined correlation measures have been obtained related to quantum channels~\cite{PhysRevA.85.032117,PhysRevA.106.012436}. Recently, the uncertainty relations of quantum channels via skew information have been also widely studied~\cite{fu2019skew,zhang2021a,sun2021uncertainty,zhang2021tighter,liu2022total,zhou2023uncertainty,li2022tighter,zhang2023note,zhang2023wigneryanase}.

In this paper, we study the uncertainty relations of channels with respect to any mixed states via variances and Wigner-Yanase skew information. The rest of the paper is arranged as follows. In Sec.~\ref{sec2}, we first review the relations between state and channel from an algebraic perspective via symmetric Jordan product and skew-symmetric Lie product, and recall their basic properties. After introducing the variance-based uncertainty for quantum channels, we define a new uncertainty to reveal better quantum mechanical nature. We establish several uncertainty relations based on the new uncertainty. Finally, we conclude in Sec.~\ref{sec4}.

\section{Uncertainty relations of quantum channels}\label{sec2}
Recall that a quantum channel $\Phi$ acting on $d$-dimension quantum state $\rho\in H_d$ can be described by the Kraus operators $\{K_i\}_{i=1}^n$,
\begin{equation}
\Phi(\rho)=\sum_i K_i\rho K_i^\dagger=\sum_i (K_i\sqrt{\rho})(K_i\sqrt{\rho})^\dagger,
\end{equation}
where $\sum_i K_i^\dagger K_i=\mathbb{I}$ with $\mathbb{I}$ the identity operator.
As the symmetry (asymmetry) is related to commutativity (non-commutativity), one can decompose the constituent interaction $K_i\sqrt{\rho}$ into
\begin{equation*}
K_i\sqrt{\rho}=\frac{\{K_i,\sqrt{\rho}\}+[K_i,\sqrt{\rho}]}{2},
\end{equation*}
where $\{K_i,\sqrt{\rho}\}=K_i\sqrt{\rho}+\sqrt{\rho}K_i$ is the symmetric Jordan product (anti-commutator) and $\left[K_i,\sqrt{\rho}\right]=K_i\sqrt{\rho}-\sqrt{\rho}K_i$ is the skew-symmetric Lie product (commutator). To quantify the symmetry and asymmetry of the state $\rho$ with respect to the channel $\Phi$, Luo and Sun defined two quantities~\cite{luo2018coherence},
\begin{align}
I_\rho(\Phi)&=\frac{1}{2}\sum_i\| [\sqrt{\rho},K_i]\|_F^2,\\
J_\rho(\Phi)&=\frac{1}{2}\sum_i\| \{\sqrt{\rho},K_i\}\|_F^2,
\end{align}
where $\|\cdot\|_F$ denotes the Frobenius norm, $\|O\|_F =\sqrt{\mathbf{tr}(O^\dagger O})$. Denote $I_\rho(K_i)=\frac{1}{2}\| [\sqrt{\rho},K_i]\|_F^2$ and $J_\rho(K_i)=\frac{1}{2}\| \{\sqrt{\rho},K_i\}\|_F^2$. $I_\rho(K_i)$ is actually the so called Wigner-Yanase skew information of $\rho$ with respect to the operator $K_i$, while $J_\rho(K_i)$ is in some sense dual to $I_\rho(K_i)$~\cite{wigner1963information}. Note that the Kraus representations of a channel are not unique, but unitary equivalent~\cite{nielsen2002quantum}. Luo and Sun proved that $I_\rho(\Phi)$ and $J_\rho(\Phi)$ are independent of the choices of Kraus representations~\cite{luo2018coherence}. The asymmetric part $I_\rho(\Phi)$ can be regarded as a bona fide measure for coherence of $\rho$ with respect $\Phi$~\cite{PhysRevA.106.012436,fu2021quantifying,luo2017quantum}. There are some nice properties of $I_\rho(\Phi)$~\cite{luo2018coherence}:

(1) Non-negativity: $I_\rho(\Phi)\geqslant 0$ and the equality holds if and only if $[\sqrt{\rho},K_i]=0$ for all $i$.

(2) Convexity: $I_\rho(\Phi)$ is convex with respect to $\rho$.

(3) Unitary covariance: $I_{U\rho U^\dagger}(U\Phi U^\dagger)=I_\rho(\Phi)$, where $U$ denotes any unitary operator and $U\Phi U^\dagger(\rho)=\sum_i UK_i\rho (UK_i)^\dagger$.

(4) Decreasing under the partial trace: $I_{\rho^{AB}}(\Phi^A\otimes \mathcal{I})\geqslant I_{\rho^A}(\Phi^A)$, where $\mathcal{I}$ is identity channel. In particular, $I_{\rho^A\otimes \rho^B}(\Phi^A\otimes \mathcal{I})=I_{\rho^A}(\Phi^A)$, where $\rho^A$ $(\rho^B)$ is any state in Hilbert space $H^A$ $(H^B)$.

The properties of $J_\rho(\Phi)$ are similar to that of $I_\rho(\Phi)$, but in a dual fashion.
In particular, when the channel $\Phi$ is unital, i.e., $\sum_iK_i^\dagger K_i=\sum_iK_iK_i^\dagger=\mathbb{I}$, then one has the complementary relations,
$I_\rho(\Phi)+J_\rho(\Phi)=2$. Moreover, it is verified that $0\leqslant I_\rho(\Phi)\leqslant 1 \leqslant J_\rho(\Phi)\leqslant 2$.
These relations indicate that the asymmetric part (usually related to the quantumness) cannot emerge alone without the accompanied symmetric part (usually related to the classicality)~\cite{luo2018coherence}.

For any operator $K$, the variance with respect to a state $\rho$ is defined as~\cite{zhang2020quantum}
\begin{equation}
V_\rho(K)=\frac{1}{2}\mathbf{tr}[(K^\dagger K+K K^\dagger) \rho]-|\mathbf{tr}(K \rho)|^2.
\end{equation}
We define the variance of a quantum channel $\Phi$ based on its Kraus decomposition $\{K_i \}$,
\begin{equation}
V_\rho(\Phi)=\sum_i V_\rho(K_i).
\end{equation}
Since for any other Kraus decomposition $\{K^\prime_i \}$, $K^\prime_i=\sum_j U_{ij}K_j$,
\begin{align*}
\sum_i V_\rho(K_i^\prime)
&=\sum_i \frac{1}{2}\mathbf{tr}[((K_i^\prime)^\dagger K_i^\prime+K_i^\prime (K_i^\prime)^\dagger) \rho]-|\mathbf{tr}(K_i^\prime \rho)|^2\\
&=\sum_i \sum_{jj\prime} U_{ij}^*U_{ij^\prime} \{\frac{1}{2}\mathbf{tr}[(K_j^\dagger K_{j^\prime}+K_j (K_{j^\prime})^\dagger) \rho]-\mathbf{tr}(K_j \rho)^*\mathbf{tr}(K_{j^\prime} \rho)\}\\
&=\frac{1}{2}\mathbf{tr}[(K_j^\dagger K_j+K_j K_j^\dagger) \rho]-|\mathbf{tr}(K_j \rho)|^2\\
&=\sum_i V_\rho(K_j),
\end{align*}
$V_\rho(\Phi)$ is independent of the Kraus decompositions.

The variance of quantum channel $V_\rho(\Phi)$ inherits nice properties of the variance of operators, such as concavity for quantum state. It is a hybrid of both classical mixing and quantum uncertainty~\cite{luo2005heisenberg}. Recall that the quantity $I_{\rho}(\Phi)$ is related to the quantumness, the following quantity
\begin{equation}
C_\rho(\Phi)=V_\rho(\Phi)-I_{\rho}(\Phi)
\end{equation}
characterizes the classical mixing uncertainty of the channels. When the state $\rho$ is pure, we have $C_\rho(\Phi)=0$, namely, there is no mixing. Since $V_\rho(\Phi)$ is concave and $I_{\rho} (\Phi)$ is convex, the difference $C_\rho(\Phi)$ is concave. Motivated by the quantum uncertainty for observables defined in Ref.~\cite{luo2005heisenberg}, let us define a new kind of uncertainty,
\begin{equation}
Q_\rho(\Phi)=\sqrt{V_\rho(\Phi)^2-(V_\rho(\Phi)-I_{\rho}(\Phi))^2},
\end{equation}
which involves terms of more quantum mechanical nature. According to the definition of $Q_\rho(\Phi)$, one has
\begin{equation}
I_\rho(\Phi)\leqslant Q_\rho(\Phi)\leqslant {2}V_\rho(\Phi)-I_\rho(\Phi).
\end{equation}

Denote $\tilde{I}_\rho(\Phi)={I}_\rho(\Phi)=V_\rho(\Phi)-C_\rho(\Phi)$ and
$\tilde{J}_\rho(\Phi)=V_\rho(\Phi)+C_\rho(\Phi)$. Straightforward algebraic calculations give rise to
\begin{align*}
\tilde{I}_\rho(\Phi)&=\frac{1}{2} \sum_i\mathbf{tr}([\sqrt{\rho},\tilde{K_i}]^\dagger[\sqrt{\rho},\tilde{K_i}]), \\
\tilde{J}_\rho(\Phi)&=\frac{1}{2} \sum_i\mathbf{tr}(\{\sqrt{\rho},\tilde{K_i}\}^\dagger \{\sqrt{\rho},\tilde{K_i}\}),
\end{align*}
where $\tilde{K_i}=K_i-\mathbf{tr}(K_i\rho)$. It is easily verified that
$Q_\rho(\Phi)=\sqrt{\tilde{I}_\rho(\Phi)\tilde{J}_\rho(\Phi)}$.

Define two supermatrices $K_I=([\sqrt{\rho},\tilde{K}_1]^T,[\sqrt{\rho}, \tilde{K}_2]^T,\cdots, [\sqrt{\rho},\tilde{K}_n]^T)^T$ and $K_J=(\{\sqrt{\rho},\tilde{K}_1\}^T,\{\sqrt{\rho},\tilde{K}_2\}^T,\cdots, \{\sqrt{\rho},\tilde{K}_n\}^T)^T$, where $T$ denotes the transpose of the matrix. Then
\begin{align*}
\tilde{I}_\rho(\Phi)&=\frac{1}{2}\mathbf{tr}(K_I^\dagger K_I)\equiv\frac{1}{2}\la K_I | K_I\ra,~~~\tilde{J}_\rho(\Phi)=\frac{1}{2}\mathbf{tr}(K_J^\dagger K_J)\equiv\frac{1}{2}\la K_J | K_J\ra.
\end{align*}

\begin{theorem}\label{th3}
Let $\Psi$ and $\Phi$ be two arbitrary channels on $d$-dimensional Hilbert space $H_d$ with Kraus decompositions $\Psi(\rho)=\sum_i L_i \rho L_i^\dagger$ and $\Phi(\rho)=\sum_j K_j \rho K_j^\dagger$, respectively. Then the product form-based uncertainty relation holds:
\begin{equation}\label{th3eq}
Q_\rho(\Psi)Q_\rho(\Phi)\geqslant\frac{1}{4}\sum_{ij}|\mathbf{tr} ([L_i,K^\dagger_j]\rho)|^2.
\end{equation}
\end{theorem}

{\textit{Proof.}}
Following the Cauchy-Schwarz inequality, we obtain
\begin{align*}
 \tilde{I}_\rho(\Psi)\tilde{J}_\rho(\Phi)&=\sum_{ij} I_{\rho} (\tilde{L}_i)J_{\rho} (\tilde{K}_j)\\
 &=\frac{1}{4}\sum_{ij} \la [\sqrt{\rho},\tilde{L}_i] |[\sqrt{\rho},\tilde{L}_i]\ra \la \{\sqrt{\rho},\tilde{K}_j\}|\{\sqrt{\rho},\tilde{K}_j\}\ra\\
  &\geqslant \frac{1}{4}\sum_{ij}|\la [\sqrt{\rho},\tilde{L}_i] | \{\sqrt{\rho},\tilde{K}_j\}\ra|^2\\
  &=\frac{1}{4}\sum_{ij}|\mathbf{tr} ([L_i,K^\dagger_j]\rho)|^2.
\end{align*}
Symmetrically, we get
\begin{equation*}
\tilde{I}_\rho(\Phi)\tilde{J}_\rho(\Psi)\geqslant\frac{1}{4}\sum_{ij}|\mathbf{tr} ([L_i,K^\dagger_j]\rho)|^2.
\end{equation*}
Thus we complete the proof by multiplying the above inequalities. $\Box$

In particular, for two unitary channels $U$ and $V$ on Hilbert space $H_d$ such that $U(\rho)=U\rho U^\dagger$ and $V(\rho)=V\rho V^\dagger$, Theorem \ref{th3} implies the following corollary.

\begin{corollary}
Let $U$ and $V$ be unitary channels. The following uncertainty relation holds:
\begin{equation}
Q_\rho(U)Q_\rho(V)\geqslant\frac{1}{4}|\mathbf{tr} ([U,V^\dagger]\rho)|^2.
\end{equation}
\end{corollary}

{\it Remark.} Our Theorem \ref{th3} for two channels can be generalized to the case of three channels. Let $\Psi$, $\Phi$ and $\Gamma$ be three arbitrary channels on $d$-dimensional Hilbert space $H_d$ with Kraus decompositions $\Psi=\sum_i L_i \rho L_i^\dagger$, $\Phi=\sum_j K_j \rho K_j^\dagger$ and $\Gamma=\sum_k M_k \rho M_k^\dagger$, respectively.
Following the Theorem \ref{th3}, we have
\begin{equation}\label{threec}
\begin{aligned}
&Q_\rho(\Psi)Q_\rho(\Phi)\geqslant\frac{1}{4}\sum_{ij}|\mathbf{tr} ([L_i,K^\dagger_j]\rho)|^2,\\
&Q_\rho(\Psi)Q_\rho(\Gamma)\geqslant\frac{1}{4}\sum_{ik}|\mathbf{tr} ([L_i,M^\dagger_k]\rho)|^2,\\
&Q_\rho(\Gamma)Q_\rho(\Phi)\geqslant\frac{1}{4}\sum_{kj}|\mathbf{tr} ([M_k,K^\dagger_j]\rho)|^2.
\end{aligned}
\end{equation}
The above relations may give rise to the following uncertainty relation,
\begin{equation*}
Q_\rho(\Psi)Q_\rho(\Phi)Q_\rho(\Gamma)\geqslant \frac{1}{8}\left\{ \sum_{ij}|\mathbf{tr} ([L_i,K^\dagger_j]\rho)|^2\sum_{ik}|\mathbf{tr} ([L_i,M^\dagger_k]\rho)|^2\sum_{kj}|\mathbf{tr} ([M_k,K^\dagger_j]\rho)|^2\right\}^\frac{1}{2}.
\end{equation*}
However, the equality may not hold since equalities in (\ref{threec}) may not be hold simultaneously. Motivated by \cite{PhysRevLett.118.180402}, the uncertainty inequality can be tighten by multiply a constant $\tau$ which is larger than $1$,
\begin{equation*}
Q_\rho(\Psi)Q_\rho(\Phi)Q_\rho(\Gamma)\geqslant \frac{\tau}{8}\left\{ \sum_{ij}|\mathbf{tr} ([L_i,K^\dagger_j]\rho)|^2\sum_{ik}|\mathbf{tr} ([L_i,M^\dagger_k]\rho)|^2\sum_{kj}|\mathbf{tr} ([M_k,K^\dagger_j]\rho)|^2\right\}^\frac{1}{2},
\end{equation*}
where $\tau$ is the minimal value running over all quantum states such that the equality holds. For the unitary channels given by the three standard Pauli matrices $\sigma_x$, $\sigma_y$ and $\sigma_z$, we have the following tight uncertainty relation,
\begin{equation}\label{tpc}
Q_\rho(\sigma_x)Q_\rho(\sigma_y)Q_\rho(\sigma_z)\geqslant\frac{\tau}{8} |\mathbf{tr} (\sigma_x\rho)\mathbf{tr} (\sigma_y\rho)\mathbf{tr} (\sigma_z\rho)|,
\end{equation}
where $\tau=\frac{64}{3\sqrt{3}}$. Let $r_x$, $r_y$, and $r_z$ be the components of the Bloch vector $\vec{r}$ of a qubit state with the density matrix $\rho=\frac{1}{2}(\mathbb{I}+\vec{r}\cdot\vec{\sigma})$. The equality in (\ref{tpc}) holds when $|r_x|=|r_y|=|r_z|=\frac{1}{\sqrt{3}}$ or $\left(|r_x|-1\right)\left(|r_y|-1\right)\left(|r_z|-1\right)=0$.

Instead of the product form uncertainty relations, the summation form uncertainty relations provides another way to capture the incompatibility of quantum observables \cite{PhysRevA.77.022105,PhysRevLett.113.260401}. Below is the summation form uncertainty relation for quantum channels.

\begin{theorem}\label{th4}
Let $\Psi$ and $\Phi$ be two arbitrary channels on $d$-dimensional Hilbert space $H_d$ with Kraus decompositions $\Psi=\sum_i L_i \rho L_i^\dagger$ and $\Phi=\sum_j K_j \rho K_j^\dagger$, respectively. Then the following summation form uncertainty relation holds,
\begin{equation}\label{th3eq}
Q^2_\rho(\Psi)+Q^2_\rho(\Phi)\geqslant\frac{1}{2}\sum_{ij}|\la [\sqrt{\rho},{L}_i] | [\sqrt{\rho},{K}_i]\ra\left(\la \{\sqrt{\rho},{L}_j\} | \{\sqrt{\rho},{K}_j\}\ra-4\la L_j^\dagger\ra\la K_j\ra\right)|.
\end{equation}
\end{theorem}
 {\textit{Proof.}}
Following the geometric-arithmetic mean inequality, we obtain
\begin{align*}
Q^2_\rho(\Psi)+Q^2_\rho(\Phi)=&\tilde{I}_\rho(\Psi)\tilde{J}_\rho(\Psi)+\tilde{I}_\rho(\Phi)\tilde{J}_\rho(\Phi)\\
=&\sum_{ij} \left[ I_{\rho} (\tilde{L}_i)J_{\rho} (\tilde{L}_j)+ I_{\rho} (\tilde{K}_i)J_{\rho} (\tilde{K}_j)\right]\\
=&\frac{1}{4}\sum_{ij} \la [\sqrt{\rho},\tilde{L}_i] |[\sqrt{\rho},\tilde{L}_i]\ra \la \{\sqrt{\rho},\tilde{L}_j\}|\{\sqrt{\rho},\tilde{L}_j\}\ra\\
 &+ \la [\sqrt{\rho},\tilde{K}_i] |[\sqrt{\rho},\tilde{K}_i]\ra \la \{\sqrt{\rho},\tilde{K}_j\}|\{\sqrt{\rho},\tilde{K}_j\}\ra\\
\geqslant &\frac{1}{2}\sum_{ij}|\la [\sqrt{\rho},\tilde{L}_i] | [\sqrt{\rho},\tilde{K}_i]\ra\la \{\sqrt{\rho},\tilde{L}_j\} | \{\sqrt{\rho},\tilde{K}_j\}\ra|\\
  =&\frac{1}{2}\sum_{ij}|\la [\sqrt{\rho},{L}_i] | [\sqrt{\rho},{K}_i]\ra\left(\la \{\sqrt{\rho},{L}_j\} | \{\sqrt{\rho},{K}_j\}\ra-4\la L_j^\dagger\ra\la K_j\ra\right)|.
\end{align*}
Thus we complete the proof. $\Box$

In particular, for unitary quantum channels we have

\begin{corollary}
Let $U$ and $V$ be unitary channels. The following summation form uncertainty relation holds,
\begin{equation}
Q^2_\rho(U)+Q^2_\rho(V)\geqslant \frac{1}{2} |\la [\sqrt{\rho},U] | [\sqrt{\rho},V]\ra\left(\la \{\sqrt{\rho},U\} | \{\sqrt{\rho},V\}\ra-4\la U^\dagger\ra\la V\ra\right)|.
\end{equation}
\end{corollary}

\begin{figure}[htbp]
 \centering
 \subfigure[]
 {
 \label{subfig:a} 
 \includegraphics[width=7cm]{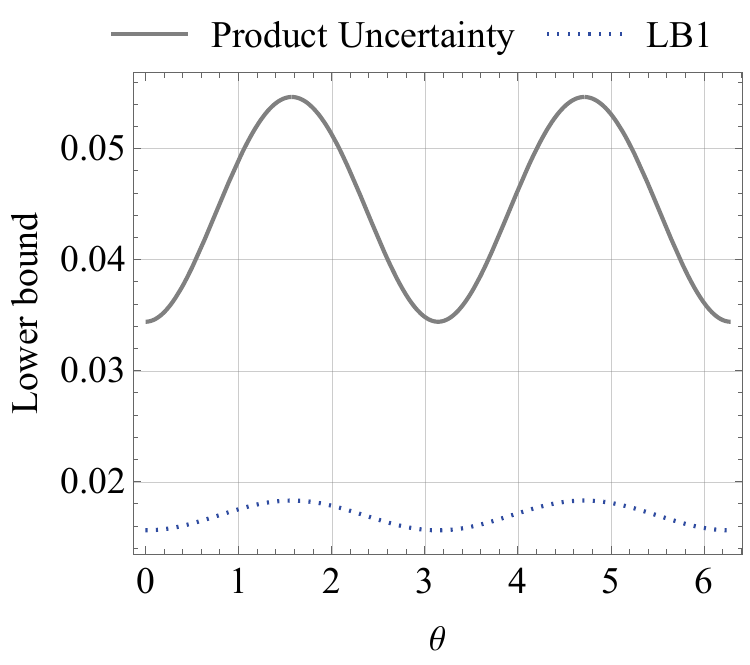}}
 \subfigure[]
 {
 \label{subfig:b} 
 \includegraphics[width=7cm]{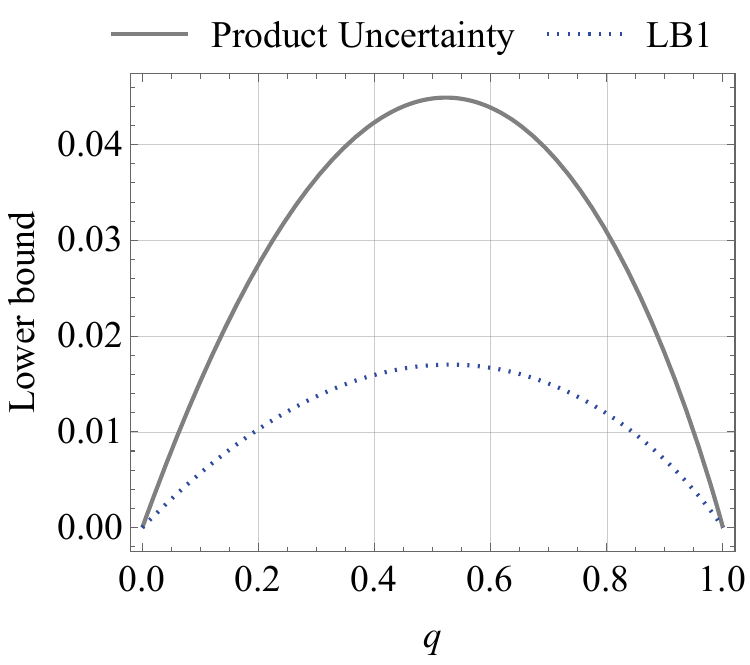}}
  \subfigure[]
 {
 \label{subfig:c} 
 \includegraphics[width=7cm]{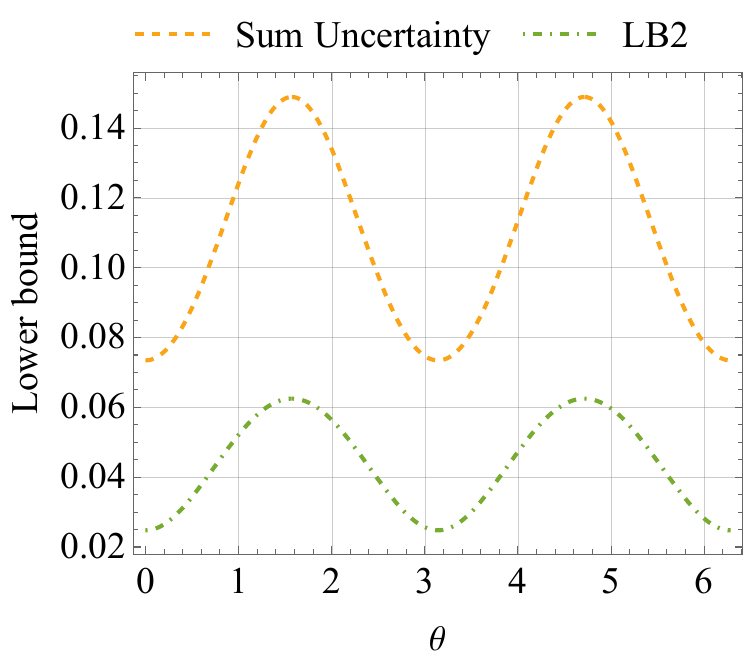}}
 \subfigure[]
 {
 \label{subfig:d} 
 \includegraphics[width=7cm]{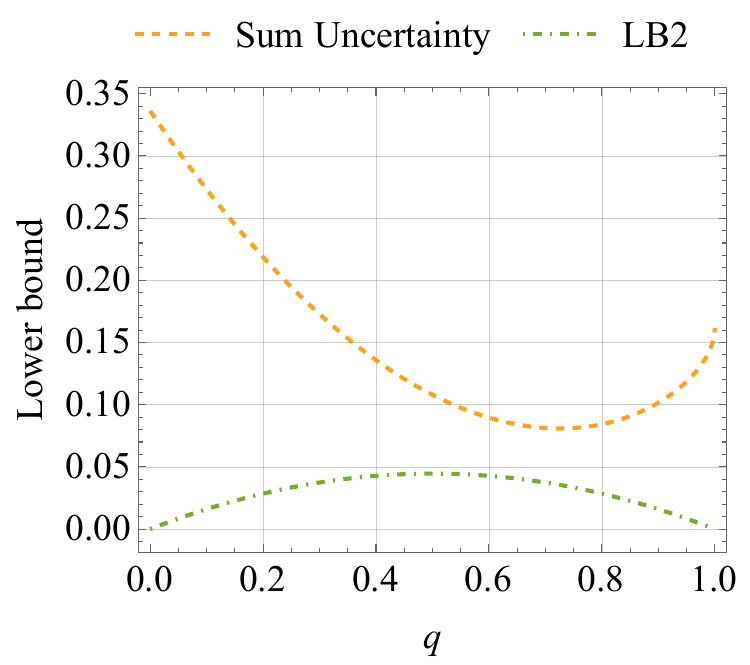}}
 \caption{The grey (solid) curve and the yellow (dashed) curve respectively represent product and sum uncertainties  in the mixed state $\rho=\frac{1}{2}(I+\vec{r}\cdot\vec{\sigma})$ with $\vec{r}=(\frac{1}{2}\cos\theta,\frac{{1}}{2}\sin\theta,\frac{1}{2})$. The blue (dotted) curve and the green (dot-dashed) curve represent our lower bounds of Theorems \ref{th3} and \ref{th4}, respectively. The comparisons of our uncertainty relations for the amplitude damping channel $\epsilon$ and the bit flip channel $\Lambda$ with $q=0.5$ (Fig.~\ref{subfig:a} and Fig.~\ref{subfig:c}) and $\theta=\pi/4$ (Fig.~\ref{subfig:b} and Fig.~\ref{subfig:d}). }
 \label{fig2}
 \end{figure}

We have established product and summation form uncertainty relations for quantum channels. For convenience, we denote the lower bounds of the uncertainty relations of Theorem \ref{th3}, and \ref{th4} by {\rm LB1} and {\rm LB2}, respectively. Let us employ an example to illustrate these lower bounds. Consider the mixed state,
\begin{equation}
\rho=\frac{1}{2}(\mathbb{I}+\vec{r}\cdot\vec{\sigma}),
\end{equation}
where $\vec{r}=(\frac{{1}}{2}\cos\theta,\frac{{1}}{2}\sin\theta,\frac{1}{2})$, and $\sigma_x$, $\sigma_y$ and $\sigma_z$ are Pauli matrices. We respectively consider two quantum channels: the amplitude damping channel $\epsilon(\rho)=\sum_{i=1}^2L_i\rho (L_i)^\dagger$, where
\begin{equation*}
\begin{gathered}
L_1=
\begin{pmatrix} 1  & 0 \\ 0  & \sqrt{1-q}  \end{pmatrix},~~~
L_2=
\begin{pmatrix} 0  & \sqrt{q} \\ 0  & 0  \end{pmatrix},
\end{gathered}
\end{equation*}
and the bit flip channel $\Lambda(\rho)=\sum_{i=1}^2K_i\rho (K_i)^\dagger$, where
\begin{equation*}
\begin{gathered}
K_1=
\begin{pmatrix} \sqrt{q}  & 0 \\ 0  &\sqrt{q}   \end{pmatrix},~~~
K_2=
\begin{pmatrix} 0  & \sqrt{1-q} \\ \sqrt{1-q}  & 0  \end{pmatrix},~~~ 0\leq q<1.
\end{gathered}
\end{equation*}
The results are shown in Fig. \ref{fig2}.

\section{Conclusion}\label{sec4}
We have established product and sum uncertainty relations for quantum channels via variances and the Wigner-Yanase skew information. These results generalize the uncertainty relations for quantum observables to quantum channels. The new defined uncertainty may reveal more quantum features comparing with the uncertainty only based on Wigner-Yanase skew information. We have considered two typical channels: amplitude damping channel and bit flip channel to illustrate the performance of these uncertainty relations. Remarkably, it is worth to mention that here we considered the Wigner-Yanase skew information to define the uncertainty. In fact, our approach may be also applied to use Wigner-Yanase-Dyson skew information, Fisher information and metric-adjusted skew information to investigate the uncertainties of quantum channels.

\bigskip
\noindent{\bf Acknowledgments}\, \,
This work is supported by the National Natural Science Foundation of China (NSFC) under Grants 12075159 and 12171044, Beijing Natural Science Foundation (Grant No. Z190005), Academician Innovation Platform of Hainan Province, and Changsha University of Science and Technology (Grant No. 097000303923).

\noindent{\bf Author Contributions}\, \,
The first author wrote the main manuscript text, and all authors reviewed and edited the manuscript.

\noindent{\bf Data Availability}\, \,
All data generated or analyzed during this study are included in the article.

\noindent{\bf Declarations}\, \,
The authors declare no competing interests.

\bibliographystyle{apsrev4-2}
\bibliography{zhang_scinter}

\end{document}